\documentclass[12pt,preprint]{aastex}
\usepackage{psfig}

\newcommand{\Wo}{\mbox{${\rm W}_0$}}
\newcommand{\msun}{\mbox{${\rm M}_\odot$}}

\newcommand{\msunppc}{\mbox{${\rm M}_\odot\,{\rm pc}^{-3}$}}

\newcommand{\nbody}{\mbox{{{\em N}-body}}}

\newcommand{\thc}{\mbox{${t_{\rm hm}}$}}

\newcommand{\trlx}{\mbox{$t_{\rm rlx}$}}

\newcommand{\tdf}{\mbox{${t_{\rm df}}$}}
\newcommand{\tcc}{\mbox{${t_{\rm cc}}$}}

\newcommand{\mm}{\mbox{$\langle m \rangle$}}

\newcommand{\rc}{\mbox{$r_{\rm core}$}}
\newcommand{\rvir}{\mbox{$r_{\rm vir}$}}

\newcommand{\rhm}{\mbox{$r_{\rm hm}$}}

\newcommand{\rtide}{\mbox{$r_{\rm tide}$}}

\newcommand{\lnl}{\mbox{$\ln \lambda$}}
\newcommand{\lnL}{\mbox{$\ln \Lambda$}}

\def\apgt{\ {\raise-.5ex\hbox{$\buildrel>\over\sim$}}\ }
\def\aplt{\ {\raise-.5ex\hbox{$\buildrel<\over\sim$}}\ }

\begin{document}


\title{The origin of IRS\,16: dynamically driven inspiral of a dense
	star cluster to the Galactic center?}

\author{Simon F.\ Portegies Zwart,\altaffilmark{1, 2}
        Stephen L.\ W.\ McMillan,\altaffilmark{3}
        Ortwin Gerhard\altaffilmark{4}
}

\altaffiltext{1}{Astronomical Institute 'Anton Pannekoek', 
                Univeristy of Amsterdam, Kruislaan 403}
\altaffiltext{2}{Section Computational Science,
                 Univeristy of Amsterdam, Kruislaan 403}
\altaffiltext{3}{Department of Physics,
                 Drexel University,
                 Philadelphia, PA 19104, USA}
\altaffiltext{4}{Astronomical Institute of the University of Basel,
                 Venusstrasse 7, CH-4102 Binningen, Switzerland}

\lefthead{Portegies Zwart et al.}

\begin{abstract}
We use direct N-body simulations to 
study the inspiral and internal evolution of dense star clusters
near the Galactic center.  These clusters sink toward the center due
to dynamical friction with the stellar background, and may go into
core collapse before being disrupted by the Galactic tidal field.  If
a cluster reaches core collapse before disruption, its dense core,
which has become rich in massive stars, survives to reach close to
the Galactic center.  When it eventually dissolves, the cluster
deposits a disproportionate number of massive stars in the innermost
parsec of the Galactic nucleus.  Comparing the spatial distribution
and kinematics of the massive stars with observations of 
IRS\,16, a group of young He I stars near the Galactic center, 
we argue that this association may have formed in this way.
\end{abstract}


\section{Introduction}

Krabbe et al.\,(1995) found $\sim 15$ bright He I emission line stars
within about 1 pc of the Galactic center, accompanied by many less
luminous stars of spectral types O and B (Genzel et
al. 2000).\nocite{2000MNRAS.317..348G} Genzel et al.~(2000) have
measured accurate positions and velocities of 41 early type stars in
this region, and report proper motions for 26 of them.  These stars
are part of the co-moving group IRS\,16, which was apparently formed
7--8\,Myr ago in a starburst of mass $\apgt 10^4$\,{\msun} (Tamblyn \&
Rieke 1993).\nocite{1993ApJ...414..573T} They show a high degree of
anisotropy; most of the He I stars in the Galactic center are on
tangential orbits (Genzel et al.\,2000).  Detailed spectroscopic
analysis of these Galactic center objects (Najarro et
al.~1994)\nocite{1994A&A...285..573N} indicates that they are highly
evolved, with a high surface ratio of helium to hydrogen $n_{\rm
He}/n_{\rm H} = 1$--1.67.  Allen et
al.\,(1990)\nocite{1990MNRAS.244..706A} classify them as Ofpe/WN9
stars, while Najarro et al.\,(1997)\nocite{1997A&A...325..700N}
identify them as luminous blue variables (LBVs) with masses between 60
and 100\,\msun.

LBVs are the late evolutionary stages of very massive ($\apgt
40$\,{\msun}) stars (Langer et al.\,1994).\nocite{1994A&A...290..819L}
Massive stars remain in this stage for only a short while ($\sim
3\times10^4$ years) after leaving the main sequence and the
helium-rich WN stage, placing these stars in a very narrow age
bracket: 3.2--3.6\,Myr (Langer et al.~1994).  If these objects are
lower-mass (25--40\,\msun) Wolf-Rayet stars, they may be somewhat
older (5 to 7 Myr, see Testor et
al.\,1993),\nocite{1993A&A...280..426T} which is consistent with the
age estimate of 7--8 Myr obtained from the model calculations of
Tamblyn \& Rieke (1993) and the independent age determination of 3--7
Myr by Krabbe et al.\,(1995).  In either case, a firm age limit of
$\sim7$\,Myr is indicated for IRS\,16.  The absence of detectable
X-ray emission from these stars (Baganoff et al.~2001a,
2001b)\nocite{2001Natur.413...45B}\nocite{2001AAS...198.8708B} argues
in favor of the LBV interpretation, in which case the age limit drops
to $\sim3.5$\,Myr.

While the age of the IRS\,16 group is fairly well constrained, the
location at which it formed is not.  One obvious possibility is that
the starburst occurred at roughly the Galactocentric radius where the
group is now observed.  However, this model is problematic, as the
formation of stars within a parsec of the Galactic Center is
difficult.  The tidal field of the central black hole is sufficient to
unbind gas clouds with densities $\aplt 10^{10}$\,cm$^{-3}$ (Morris
1993). At a distance $\apgt 1$\,pc star formation is still easily
prevented, even though the potential of the bulge starts to dominate
over that of the black hole.

Gerhard (2001)\nocite{2001ApJ...546L..39G} proposed that a
massive star cluster of mass $m$ formed at a distance of 
$\aplt 30 (m/10^6$\,\msun)\,pc 
from the Galactic center can spiral in to the Galactic center by
dynamical friction before being disrupted by the tidal field of the
Galaxy or its own internal evolution.  In order to survive in the
Galactic central region the cluster core density has to exceed $\rho_c
\apgt 10^7$\,\msunppc.  It is unlikely that a star cluster would be
born
with such a high central density, but it may evolve into this state
when core collapse occurs.  However, even then the cluster must have
been initially quite compact.  Core collapse of a cluster boosts the
central density, but can be strongly affected by mass loss from the
cluster tidal boundary.  In the strong tidal environment of the
Galactic center, mass loss from the cluster perimeter may prevent core
collapse altogether.

We simulate dense star clusters using a direct {\nbody} approach,
taking the external potential of the Galaxy and the effect of
dynamical friction into account.  Within this model we study
the possibility that a cluster may go into core collapse before
dynamical friction causes it to spiral in to the Galactic center.  We
include the dynamical friction term analytically, applying it to the
bound cluster mass (see Binney \& Tremaine
1987).\nocite{1987gady.book.....B} In \S2 we discuss cluster evolution
and dynamical friction, in order to interpret the results of our model
calculations, which are presented in \S\ref{Sect:cluster_survival}.
We summarize in \S\ref{Sect:Summary}.

\section{Cluster dynamics}

\subsection{Time scale for core collapse}

The dynamical evolution of a star cluster drives it toward core
collapse (Antonov 1962; Spitzer \& Hart, 1971)
\nocite{1962spss.book.....A}\nocite{1971ApJ...164..399S} in which the
central density runs away to a formally infinite value in a finite
time.  Core collapse occurs at
\begin{equation}
        \tcc \simeq c \,\trlx\,,
\label{Eq:tcc}\end{equation}
where {\trlx} is the cluster's ``half-mass'' relaxation time,
\begin{equation}
        \trlx = {\rvir^{3/2} \over (Gm)^{1/2}}
                {n \over 8 \lnl}.
\end{equation}
Here $G$ is the gravitational constant, $n$ is the number of stars in
the cluster, $m$ is the total cluster mass, and {\rvir} is the
cluster's virial radius.  The Coulomb logarithm $\lnl \simeq \ln (0.1
n) \sim 10$ typically.

In an isolated cluster in which all stars have the same mass, $c
\simeq 15$ (Cohn 1980).
\nocite{1980ApJ...242..765C}\nocite{2001A&A...375..711F} In a
multi-mass system, core collapse is determined by the accumulation of
the most massive stars in the cluster center (Vishniac 1978; see also
Chernoff \& Weinberg
1990)\nocite{1990ApJ...351..121C}\nocite{1978ApJ...223..986V}.  We
have performed direct {\nbody} simulations to determine the moment at
which core collapse occurs, and hence the value of $c$ in
Eq.\,\ref{Eq:tcc}.

The initial conditions of our model cluster are presented in
Table\,\ref{Tab:isolated}.  The cluster consists of 65536 stars
distributed initially in a King (1966) model with King parameter $\Wo
= 3$.  Each of the stars is randomly assigned a mass drawn from a
Scalo (1986) initial mass function between 0.3 and 100\,\msun,
irrespective of position.  The entire cluster is then rescaled to
virial equilibrium.  We choose a virial radius $\rvir = 0.167$\,pc.
These choices mimic the young Arches and Quintuplet star clusters,
which are located somewhat farther ($\sim30$\,pc) from the Galactic
center.  The resulting initial parameters (total mass, core radius,
half mass radius, crossing time and relaxation time) are also listed
in the table.

\begin{table*}[htbp!]
\caption[]{Initial conditions for the simulated star clusters.  The
columns give the number of stars, mass (in \msun), the King parameter
{\Wo}, the core-, half mass-, virial- and tidal radii (all in
parsecs).  The last two columns give the initial half mass crossing
time and the two-body relaxation (both in millions of years).}
\begin{flushleft}
\begin{tabular}{lrrrrrrrrrrrrr} \hline
\hline                         
$n$&  $m$& \Wo&   \rc&  \rhm& \rvir& \rtide& \thc& \trlx \\
   &[\msun]&  & \multicolumn{4}{c}{------------ [pc] ------------}
                  & \multicolumn{2}{c}{--- [Myr] ---} \\
\hline
65536&62549&   3& 0.116& 0.140& 0.167& 0.525& 0.0115&  3.87 \\
\hline
\end{tabular}
\end{flushleft}
\label{Tab:isolated}
\end{table*}

Visual inspection of the core radius as a function of time indicates
that core collapse occurs around $t=0.76$\,Myr, near the moment when
the hard binary containing the most massive star reaches a binding
energy of $E < -100\,kT$ (where $kT$ is the thermodynamic energy scale
of the stellar system; the total kinetic energy of the cluster is
$\frac{3}{2} nkT$).  This binary was formed somewhat earlier (at $t =
0.58$\,Myr), but at that time we could not identify the core as
collapsed, as the core radius continued to contract.  A little later
($t=0.84$\,Myr), this binary is strongly perturbed by another star,
resulting in a collision.  Based on this information, we conclude that
this particular simulation experienced core collapse at
$t\sim0.76$\,Myr, so $c \simeq 0.20$, which is consistent with
Portegies Zwart \& McMillan (2002),\nocite{2002astro.ph..1055P} but
somewhat larger than the $c \simeq 0.15$ found in the Fokker-Planck
simulations of Chernoff \& Weinberg
(1990).\nocite{1990ApJ...351..121C}

\subsection{Inspiral to the Galactic center}

The mass $M$ of the Galaxy within the cluster's orbit at distance $R$
($\aplt500$\,pc) from the Galactic center is taken to be (Sanders \&
Lowinger 1972; Mezger et
al.\,1996)\nocite{1996A&ARv...7..289M}\nocite{1972AJ.....77..292S}
\begin{equation}
        M(R) = AR^\alpha,
\label{Eq:Galaxymass}\end{equation}
where $A=4.25\times10^6\,\msun (1 {\rm pc})^{-\alpha}$ and
$\alpha=1.2$. This slope fits the observed light distribution with
constant $M/L$ and the rotation curve derived from OH/IR stars and
21cm line observations (Mezger et al.\,1996). Earlier observations,
however, claim a slightly shallower slope (Catchpole, Whitelock \&
Glass 1990).  For clarity we adopt $\alpha = 1.2$ for the remainder of
this paper.  The density at distance $R$ then is
\begin{equation}
        \rho(R) = {1 \over 4\pi R^2} {dM \over dR}
                = {A\alpha \over 4\pi} R^{\alpha-3}.
\label{Eq:rho}\end{equation}
Following Binney \& Tremaine (1987), we find that the inspiral of the
cluster towards the center due to dynamical friction is described by
(McMillan \& Portegies Zwart, 2002, ApJ in press)
\begin{equation}
        {dR \over dt} = - 2\ln \Lambda {\alpha \chi \over \alpha+1}
                        \left({G \over A}\right)^{1/2}
                         m R^{-\frac12 (\alpha+1)}.
\label{Eq:dRdt}\end{equation}
Here $m$ is constant and $\ln\Lambda \sim \ln(R/\langle r\rangle) \sim
5$ is a Coulomb logarithm (where $\langle r\rangle$ is the object's
characteristic radius---roughly the half-mass radius in the case of a
cluster) and $\chi\sim0.3$ is a parameter which depends on the
velocity of the cluster and the velocity dispersion of the stellar
surroundings.  In this case, $\chi\ln\Lambda\sim1$.  The adopted value
of $\ln \Lambda$ is consistent with results from N-body simulations
(Spinnato et al.\, 2003), who derive $\ln \Lambda = 6.6\pm0.6$ for a
massive compact object that spirals-in, and somewhat smaller than the
value $\ln \Lambda \simeq \ln 0.4N$ used by Gerhard (2001), where $N$
the number of stars with which the cluster interacts. For simplicity
we write Eq.\,\ref{Eq:dRdt} as
\begin{equation}
        R^{\frac12 (\alpha+1)} dR = -\gamma dt.
\end{equation}

Solving the differential equation Eq.\,\ref{Eq:dRdt} with $R(0) = R_i$
at time $t = 0$ results in
\begin{equation}
        R(t) = R_i \left[
                        1 - {(3+\alpha)\gamma \over 2R_i^{\frac12 (3+\alpha)}}
                        t
                   \right]^{2/(3+\alpha)}.
\end{equation}
Inverting this equation with $R = R_f$ (the disruption radius) at
$t=\tdf$ and substituting 
Eq.\,\ref{Eq:Galaxymass} gives
\begin{equation}
        \tdf = {\alpha+1 \over \alpha(\alpha+3)} {1 \over \chi \ln \Lambda}
                m^{-1}
                \left( {M(R_i) \over G} \right)^{1/2}
                \left [R_i^{3/2} - \kappa R_f^{3/2} \right],
\label{Eq:tdf}\end{equation}
where $\kappa = (R_f/R_i)^{\alpha/2}$.  For $\alpha=1.2$,
$\chi\ln\Lambda=1$, and $R_f \ll R_i$, this becomes
\begin{equation}
        \tdf = 1.34 \left(m\over10^4\msun\right)^{-1} \left(R_i\over1
                   {\rm pc}\right)^{(3+\alpha)/2} {\rm Myr}\,.
\end{equation}

\section{Results}\label{Sect:cluster_survival}

In order to test the hypothesis that a cluster can experience core
collapse before reaching the Galactic center, it is instructive to
compare the dynamical friction inspiral time scale with the time scale
for internal cluster evolution.  We define
\begin{eqnarray}
        \eta &\equiv& {\tcc \over \tdf} \nonumber \\
               &\simeq& {\alpha (\alpha+3) \over \alpha+1}
                        {c \chi \lnL \over 8 \lnl}
                        \left( {m \over M } \right)^{1/2}
                        {m \over \mm}
                        {\rvir^{3/2} \over R_i^{3/2} - \kappa R_f^{3/2}}\,.
\end{eqnarray}
Here \mm\, is the mean mass of cluster stars.  For small $R_f$, this
reduces to
\begin{equation}
        \eta \simeq \left({0.29 c \chi \lnL\over \lnl}\right)\,
                        n\,
                        \left( {m \over M } \right)^{1/2}
                        \left( {\rvir \over R_i} \right)^{3/2}\,.
\label{Eq:alpha}\end{equation}

There are now three distinct regimes: (i) If $\eta\ll 1$ (far from
the Galactic center: $[R/{\rm pc}]^{4.2} \gg [n/2.1\times10^3]^2\,
[m/\msun]\,[\rvir/{\rm pc}]^{3}$), the cluster core collapses
essentially at its original distance from the Galactic center;
thereafter it dissolves, mainly by tidal stripping and mass loss due
to stellar evolution, at constant Galactocentric radius.  (ii) If
$\eta\sim1$ (intermediate distance to the Galactic center), cluster
inspiral and core collapse occur on about the same time scale.  (iii)
If $\eta\gg 1$ (close to the Galactic center), the cluster spirals
in without significant internal evolution.

For example, substituting the initial conditions of
Tab.\,\ref{Tab:isolated} ($m=65\,000$\,\msun, $\rvir = 0.167$\,pc) and
Eq.\,\ref{Eq:Galaxymass} into Eq.\,\ref{Eq:alpha}, we can write
\begin{equation}
        \eta \simeq 150 c \chi {\lnL \over \lnl} R^{-2.1}.
\end{equation}
Taking $c \chi \lnL/\lnl \sim 0.13$, we find that this cluster will
experience core collapse before it reaches the Galactic center if it
was born at $R_i \apgt 4$\,pc.

\begin{figure}[htbp!]
\psfig{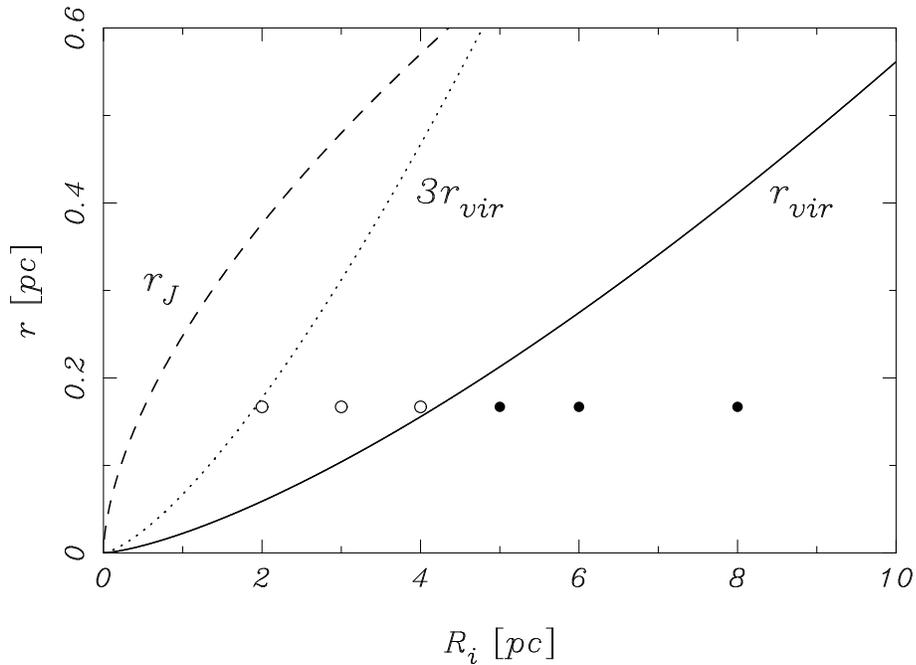}
\caption{ Virial radius {\rvir} of a cluster for which core collapse
is expected by the time of arrival in the Galactic center (solid
curve).  Here the mass of the cluster is 65\,000\,\msun, with 64k
stars. The dotted curve gives 3\,{\rvir} which corresponds to the
radius where the cluster density goes to zero for a King model with
$\Wo=3$.  The dashed line gives the Jacobi radius of the cluster in the
tidal field of the Galaxy.  Circles (lower left) and bullets show the
simulations performed for this study.  Circles indicate that the
cluster dissolved before core collapse; bullets indicate simulations
which did experience core collapse.
\label{fig:Rgc_alpha_II}
}
\end{figure}

More generally, Fig.\,\ref{fig:Rgc_alpha_II} shows, as functions of
Galactocentric radius, the virial radii (solid curve) and an estimate
for the initial tidal radii (dots) of star clusters with masses and
structure parameters as listed in Tab.\,\ref{Tab:isolated}, for
clusters with $\eta=1$ in Eq.\,\ref{Eq:alpha} (core collapse upon
arrival at the Galactic center), and take $c\chi\lnL/\lnl\sim0.13$.
The dashed curve presents an estimate of the Jacobi radius of the star
cluster in the Galactic tidal field (see Eq.\,4 in Gebhard 2001, or
Eq.\,24 in McMillan \& Portegies Zwart 2002). A 65\,000\,{\msun} star
cluster which is born with parameters to the right or below the solid
curve is expected to experience core collapse before it reaches the
Galactic center.

The circles and bullets at $\rvir = 0.167$\,pc in
Fig.\,\ref{fig:Rgc_alpha_II} indicate the outcomes of simulations
performed with cluster initial conditions as presented in
Tab.\,\ref{Tab:isolated}, but with varying values of initial
Galactocentric radius $R_i$.  The initial models were placed at $R_i =
2$, 3, 4, 5, 6, and 8\,pc.  Bullets indicate that a model experienced
core collapse before reaching the Galactic center; circles indicate
disruption before significant contraction of the cluster core.

The equations of motion of the 65536 (64k) stars in the simulations
were computed using the {\tt Starlab} software environment which
combines the {\nbody} integrator {\tt kira} and the binary evolution
package {\tt SeBa} (Portegies Zwart et
al.~2001)\nocite{2001MNRAS.321..199P}\footnote{see {\tt
http://manybody.org}}.  The Galactic tidal field and the effects of
dynamical friction were taken properly into account by solving
Eq.\,\ref{Eq:dRdt} numerically during the integration of the equations
of motion.  In these calculations, the clusters could lose mass by
tidal stripping, high velocity stellar ejections, and stellar winds.
At any moment in time we determined the total cluster mass from all
bound stars; this may slightly underestimate the friction force.  The
dynamical friction term was then applied to each of the bound stars,
but not to unbound stars.  The calculations were carried out using the
special-purpose GRAPE-6 computer (Makino et al.~1997; Makino
2001).\nocite{2001dscm.conf...87M}\nocite{1997ApJ...480..432M}

\begin{table*}[htbp!]
\caption[]{
}
\begin{flushleft}
\begin{tabular}{lrrrrrrrrrrrrr} \hline
\hline                         
$R_i$ & $t_{\rm cc}$  &$R_{\rm cc}$ & $t_{\rm diss}$ & $R_f$ \\ \hline
 [pc] & [Myr]         & [pc]&     [Myr]              & [pc]  \\
2     &  ---          & --- &     1.08               & 1.1   \\    
3     &  ---          & --- &     1.01               & 1.1   \\
4     &  ---          & --- &     0.83               & 1.3   \\
5     &  0.65         & 3.3 &     1.19               & 1.8   \\
6     &  0.68         & 4.6 &     $>1.44$            & $<2.5$\\
8     &  0.56         & 7.0 &     $>1.05$            & $<6.4$\\
\hline
\end{tabular}
\end{flushleft}
\label{Tab:models}
\end{table*}

The results of the simulations in Tab.\,\ref{Tab:models}, may be
summarized as follows.  The models with $R_i = 2$ and 3\,pc both
dissolved at $R_f = 1.1$\,pc.  They did not experience core collapse,
nor were any persistent hard binaries formed.  (For definiteness, we
take a cluster to have dissolved once the bound mass drops below
6\,000\,\msun.)  The model with $R_i= 4$\,pc dissolved at
$R_f=1.3$\,pc, but core collapse in this case is uncertain.  A few
hard ($E<-10\,kT$) binaries formed at $t=0.51$\,Myr, at a distance of
$R = 2.3$\,pc.  One of these binaries hardened to $E\aplt-50\,kT$ at
$t=0.63$\,Myr and $R=1.8$\,pc; the cluster dissolved a little later,
at $t=0.83$\,Myr.  This ambiguity is consistent with the cluster's
location close to the solid curve in Fig.\,\ref{fig:Rgc_alpha_II}.
The models with $R_i \ge 5$\,pc all experienced core collapse at
$t_{cc}\sim0.6$\,Myr.  The $R_i = 5$\,pc model dissolved at
$t=1.19$\,Myr at a galactocentric distance of $R_f =
1.8$\,pc.\footnote{An animation of this simulation is available at
{\tt http://manybody.org/starlab.html}.  See also \\ {\tt
http://www.ids.ias.edu/$\sim$starlab/animations/}.}.  The other models
were not continued to the point of dissolution.

\begin{figure}[htbp!]
\psfig{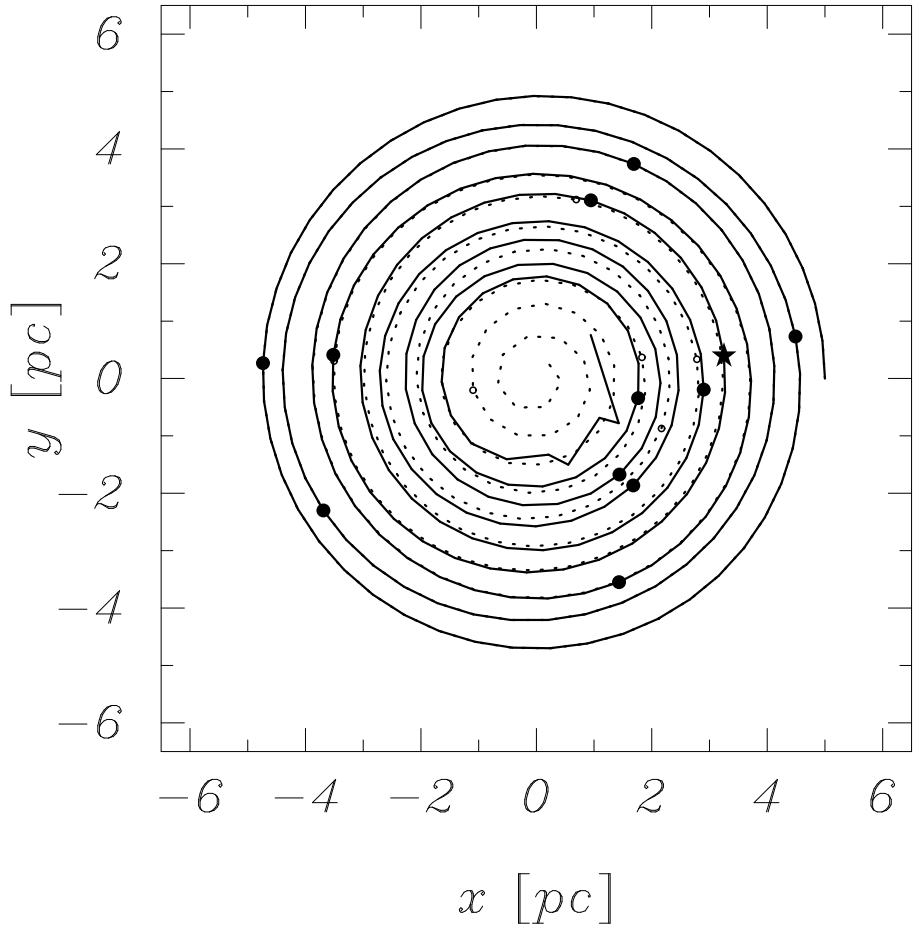}
\caption[]{Orbital evolution of the star cluster with $R_i=5$\,pc (solid
curve).  The cluster is tracked via its density center, which becomes
hard to determine accurately when the cluster contains only a few
stars, i.e: near the Galactic center.  The star indicated the moment
of core collapse. The dotted curve shows the orbit for a constant-mass
point with the same mass as the cluster at birth.  The bullets in the
orbit indicate $10^5$ year time intervals for the cluster orbit. In
the dotted curve these intervals are indicated with small circles.

\label{fig::N64R4r16XYgc}
}
\end{figure}

Figure\,\ref{fig::N64R4r16XYgc} shows a top view of the orbit of the
cluster with $R_i=5$\,pc.  The dotted curve indicates the expected
orbit of the cluster if its mass would remain constant and was not
affected by stellar evolution, internal relaxation or by the external
tidal field.  This constant-mass point spirals-in slightly more
quickly than the cluster in which mass loss is taken into account self
consistently (solid curve).

\begin{figure}
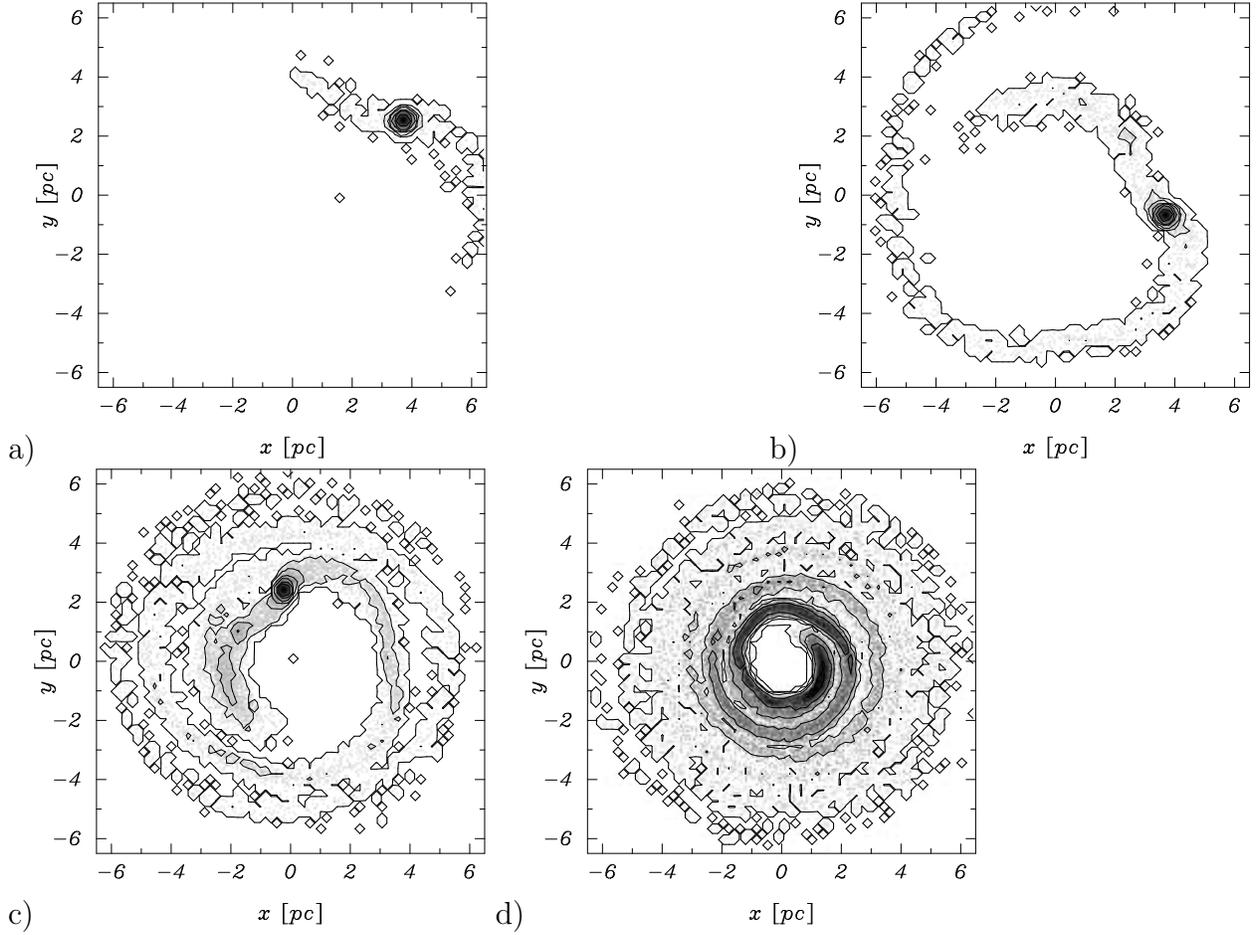
  
a)\psfig{figure=./f3a.ps,width=6.cm,angle=-90}
b)\psfig{figure=./f3b.ps,width=6.cm,angle=-90}
c)\psfig{figure=./f3c.ps,width=6.cm,angle=-90}
d)\psfig{figure=./f3d.ps,width=6.cm,angle=-90}
\caption[]{Top view of model $R_i = 5$\,pc at $t=0.1$\,Myr, 0.4\,Myr,
0.8\,Myr and 1.2\,Myr. Density is gray shaded linearly between maximum
(dark) and zero density (light) scaled individually to each panel.
The contours indicate a constant stellar density of 10 stars/pc$^3$,
50 stars/pc$^3$, 100 stars/pc$^3$, etc\footnote{see {\tt
http://manybody.org/stalab.html} for an animation}. (Note that this
calculation was performed with a different value of $\ln\Lambda$ than
was adopted in Fig.\,\ref{fig::N64R4r16XYgc}.)  }
\label{fig:XYgray_N64kW3.r0.167.R5}
\end{figure}

Figure\,\ref{fig:XYgray_N64kW3.r0.167.R5} shows four subsequent
snapshots (gray shadings and contours) at time intervals of 0.4\,Myr
for the cluster with $R_i=5$\,pc.

It is at first somewhat surprising that Eq.\,\ref{Eq:alpha} agrees so
well with the simulation results, as the cluster mass in the latter is
a function of time which is neglected in the analytic form.  Gerhard
(2001) corrects for mass loss using an isothermal model which implies
that the cluster loses mass at a constant rate until disruption. This
would result in a factor of two increase in the dynamical friction
time. It turns out that this overestimates mass loss considerably (see
fig.\,\ref{fig::N64R4r16XYgc}). Most mass in the \Wo=3 King model is
lost near the end of the cluster lifetime when the stellar density in
the environment becomes comparable to the cluster density near the
half mass radius.

In the clusters which did not experience core collapse, stars at
disruption were spread over a broad range in radii: $R_f \aplt R \aplt
R_i$ pc.  Stars more massive than 40\,{\msun} were not distributed in
a significantly different way from low-mass stars.  However, in the
$R_i=5$\,pc model, which did reach core collapse before disruption,
the massive stars became much more centrally concentrated than the
other cluster members.  The more massive stars penetrated closer to
the Galactic center because they sank to the cluster core, whereas
low-mass stars were lost from the cluster at an earlier stage, when
the cluster was farther from the center.

While we have shown that core collapsed clusters preferentially
deposit their most massive stars closest to the Galactic center, our
models differ in some important ways from the observations reported by
Genzel et al.~(2000).  The end-point of our $R_i=5$ core-collapse
cluster is a tight clump of stars with spatial extent (half-mass
radius) comparable to those observed, but the clump was deposited at a
radius of almost 2 pc, not within the innermost 1 pc, as is usually
assumed (Krabbe et al.~1995).  However, we expect that more massive
clusters (starting with smaller virial radii or larger galactocentric
radii---see Eq.\,\ref{Eq:alpha}), more centrally concentrated systems
or systems on elliptic orbits, penetrate deeper into the Galactic
potential.

Genzel et al.\,(2000) present (their Fig.\,5) the velocity anisotropy
of 12 of the He I stars near the Galactic center.  The anisotropy
parameter is $\gamma \equiv (v_t^2-v_r^2)/(v_t^2+v_r^2)$, where $v_t$
and $v_r$ are the transverse and radial velocity components of the
stellar proper motions.  The average velocity anisotropy of these
stars is $\langle \gamma \rangle = 0.59 \pm0.48$.  When we exclude the
one star with an unusually low value of $\gamma = -0.83 \pm0.33$, the
average anisotropy increases to $\langle \gamma \rangle =
0.72\pm0.18$.  We measured the velocity anisotropy among the star with
a mass $>40\,\msun$ of our $R_i=5$\,pc model at the moment it
disrupted, and found $\langle \gamma\rangle = 0.79\pm0.23$.  For all
stars in the cluster the mean velocity anisotropy is $\langle
\gamma\rangle = 0.04 \pm 0.63$, which is consistent with being
isotropic.  Likewise the sky-projected radial and tangential
velocities of all 104 proper motion stars in the sample of Genzel et
al.\,(2000) is consistent with overall isotropy.

If IRS\,16 is indeed a remnant cluster core, our simulations provide
no easy explanation of the rather broad stellar distribution
perpendicular to the supposed cluster orbit plane, nor for the large
dispersion in their velocities.  The stars in our simulations are
eventually spread out in the orbit plane, they are quite tightly
confined in the direction perpendicular to this plane.  The dispersion
in the velocity distribution of these stars then would be of the order
of the cluster velocity dispersion: of the order of ten kilometers per
second, rather than the observed dispersion of a few 100\,km/s.  The
disruption of two star clusters in short succession would not
reproduce all kinematic information.  We speculate that other
dynamical processes, such as the effects of primordial binaries, the
presence of a central black-hole binary or possible inhomogeneities in
the background potential, might have operated in IRS\,16 to increase
its scale height out of the plane and to drive the velocity dispersion
to its observed values. At present, however, we have no ready solution
to this conundrum.

\section{Summary}\label{Sect:Summary}

We have critically examined the hypothesis proposed by Gerhard (2001)
that the group IRS\,16 may be the remnant of a much larger cluster
that formed farther from the Galactic center and sank toward the
center via dynamical friction.  IRS\,16 contains about 40 early-type
stars, including at least 15 very luminous He I stars, all lying
within $\sim 0.4$\,pc of the Galactic center.

We have studied this possibility by performing a series of direct
{\nbody} simulations in which dynamical friction is taken into account
in a semi-analytic fashion, and included self-consistently in the
equations of motion of the cluster stars.  The {\nbody} calculations
were performed with 65536 stars and were run on the GRAPE-6.  Stellar
masses were selected from a realistic mass function and stars were
initially distributed as a $\Wo=3$ King model with virial radius
$\rvir = 0.167$\,pc.  We find that, in order for a clump of massive
stars to survive, the cluster must have experienced core collapse
during the inspiral.  Core collapse deposits the observed high
proportion of early type stars close to the Galactic center, and
prevents a spread of massive stars to larger distances.  The
anisotropy observed for the early-type stars in IRS\,16 is consistent
with our model calculations. However, the spatial extent and high
dispersion in the velocities of the observed cluster are not
satisfactorily explained with the current simulations. The presence of
primordial binaries or a binary of intermediate mass black holes in
the cluster center may be required to explain these observations.

Our approximation to the dynamical friction term has some limitations,
as the parameter {$\chi \lnL$} is fixed in our simulations.  In
reality, this term may differ from the adopted value, will probably
vary with time, and may depend on a number of external factors.  More
accurate measurements of this parameter are presented by Spinnato et
al.\, (2003).  Regardless of this uncertainty,
we are still able to draw some firm conclusions.  We find that dense
star clusters in a strong tidal field experience core collapse on a
time scale similar to that for an isolated cluster, and that core
collapse can occur before a cluster near the Galactic center is
tidally disrupted.

When the cluster experiences core collapse, the fraction of massive
stars deposited near the center is much greater than when core
collapse is averted by tidal disruption.  Our simulated clusters
dissolve when their core densities fall below a few million \msunppc,
more than an order of magnitude higher than the local background
stellar density.  Our calculations were performed using rather low
concentration \Wo=3 clusters.  We expect that clusters with higher
initial concentrations would penetrate deeper and more easily to the
Galactic center.  Variations in the initial orbit of the cluster may
also prove to be efficient in transporting stars closer to the
Galactic center. Note that our choice of initial conditions are
possibly among the least favorable to explain the observations.  The
parameter space for clusters which experience core collapse before
reaching the Galactic center may therefore be even larger than
suggested here.

Our model calculations support the scenario proposed by Gerhard (2001)
to explain the presence of a population of early type stars within a
parsec of the Galactic center. If born at a distance of $\sim 5$\,pc,
the primordial cloud from which the cluster formed should have had an
initial density on the order of $10^6$ cm$^{-3}$, but this density
might be lower if the stars in IRS\,16 originated in a somewhat more
massive cluster at a greater distance.

\bigskip\bigskip
\noindent{\bf Acknowledgments}

We are grateful to Rainer Spurzem for many discussions and his
hospitality, and we thank Jun Makino for the use of the GRAPE-6
systems at Tokyo University. We also thank Tokyo University, the
Institute for Advanced Study, Astronomisches Rechen Institute and the
American Museum for Natural History for their hospitality.  This work
was supported by NASA ATP grants NAG5-6964 and NAG5-9264, by the Swiss
National Science Foundation grant 20-64856.01, and by the Netherlands
Organization for Scientific Research (NWO).  SPZ was supported as
fellow of the Royal Netherlands Academy of Arts and Sciences (KNAW)
and the Netherlands Research School for Astronomy (NOVA).


\begin{thebibliography}{}

\bibitem[\protect\astroncite{{Allen} et~al.}{1990}]{1990MNRAS.244..706A}
{Allen}, D.~A., {Hyland}, A.~R., {Hillier}, D.~J. 1990, \mnras, 244, 706

\bibitem[\protect\astroncite{{Antonov}}{1962}]{1962spss.book.....A}
{Antonov}, V.~A. 1962,
\newblock {Solution of the problem of stability of stellar system Emden's
  density law and the spherical distribution of velocities},
\newblock Vestnik Leningradskogo Universiteta, Leningrad: University, 1962

\bibitem[\protect\astroncite{{Baganoff} et~al.}{2001a}]{2001Natur.413...45B}
{Baganoff}, F.~K., {Bautz}, M.~W., {Brandt}, W.~N., {Chartas}, G., {Feigelson},
  E.~D., {Garmire}, G.~P., {Maeda}, Y., {Morris}, M., {Ricker}, G.~R.,
  {Townsley}, L.~K., {Walter}, F. 2001a, \nat, 413, 45

\bibitem[\protect\astroncite{{Baganoff} et~al.}{2001b}]{2001AAS...198.8708B}
{Baganoff}, F.~K., {Bautz}, M.~W., {Ricker}, G.~R., {Brandt}, W.~N., {Chartas},
  G., {Feigelson}, E.~D., {Garmire}, G.~P., {Maeda}, Y., {Townsley}, L.~K.,
  {Morris}, M., {Walter}, F. 2001b,
\newblock in American Astronomical Society Meeting, Vol. 198, p.~8708

\bibitem[\protect\astroncite{{Binney} \&
  {Tremaine}}{1987}]{1987gady.book.....B}
{Binney}, J., {Tremaine}, S. 1987,
\newblock Galactic dynamics,
\newblock Princeton, NJ, Princeton University Press, 1987, 747 p.\,425

\bibitem[]{1990MNRAS.247..479C}
Catchpole, R.~M., Whitelock, P.~A., Glass, I.~S. 1990, MNRAS, 247, 479

\bibitem[\protect\astroncite{{Chernoff} \&
  {Weinberg}}{1990}]{1990ApJ...351..121C}
{Chernoff}, D.~F., {Weinberg}, M.~D. 1990, \apj, 351, 121

\bibitem[\protect\astroncite{{Cohn}}{1980}]{1980ApJ...242..765C}
{Cohn}, H. 1980, \apj, 242, 765

\bibitem[\protect\astroncite{{Freitag} \& {Benz}}{2001}]{2001A&A...375..711F}
{Freitag}, M., {Benz}, W. 2001, \aap, 375, 711

\bibitem[\protect\astroncite{{Genzel} et~al.}{2000}]{2000MNRAS.317..348G}
{Genzel}, R., {Pichon}, C., {Eckart}, A., {Gerhard}, O.~E., {Ott}, T. 2000,
  \mnras, 317, 348

\bibitem[\protect\astroncite{{Gerhard}}{2001}]{2001ApJ...546L..39G}
{Gerhard}, O. 2001, \apjl, 546, L39

\bibitem[]{}
King, I., 1966, AJ, 71, 276

\bibitem[]{}
Krabbe, A., Genzel, R., Eckart, A., et al., 1995, ApJ 447, L95

\bibitem[\protect\astroncite{{Langer} et~al.}{1994}]{1994A&A...290..819L}
{Langer}, N., {Hamann}, W.-R., {Lennon}, M., {Najarro}, F., {Pauldrach},
  A.~W.~A., {Puls}, J. 1994, \aap, 290, 819

\bibitem[\protect\astroncite{{Makino}}{2001}]{2001dscm.conf...87M}
{Makino}, J. 2001,
\newblock in ASP Conf. Ser. 228: Dynamics of Star Clusters and the Milky Way,
  ~87

\bibitem[\protect\astroncite{{Makino} et~al.}{1997}]{1997ApJ...480..432M}
{Makino}, J., {Taiji}, M., {Ebisuzaki}, T., {Sugimoto}, D. 1997, \apj, 480, 432

\bibitem[\protect\astroncite{{Mezger} et~al.}{1996}]{1996A&ARv...7..289M}
{Mezger}, P.~G., {Duschl}, W.~J., {Zylka}, R. 1996, \aapr, 7, 289

\bibitem[]{}
Morris, M.~1993, ApJ, 408, 496

\bibitem[\protect\astroncite{{Najarro} et~al.}{1994}]{1994A&A...285..573N}
{Najarro}, F., {Hillier}, D.~J., {Kudritzki}, R.~P., {Krabbe}, A., {Genzel},
  R., {Lutz}, D., {Drapatz}, S., {Geballe}, T.~R. 1994, \aap, 285, 573

\bibitem[\protect\astroncite{{Najarro} et~al.}{1997}]{1997A&A...325..700N}
{Najarro}, F., {Krabbe}, A., {Genzel}, R., {Lutz}, D., {Kudritzki}, R.~P.,
  {Hillier}, D.~J. 1997, \aap, 325, 700

\bibitem[\protect\astroncite{{Portegies Zwart} \&
  {McMillan}}{2002}]{2002astro.ph..1055P}
{Portegies Zwart}, S.~F., {McMillan}, S.~L.~W. 2002, ApJ, 576, 899

\bibitem[\protect\astroncite{{Portegies Zwart}
  et~al.}{2001}]{2001MNRAS.321..199P}
{Portegies Zwart}, S.~F., {McMillan}, S. L.~W., {Hut}, P., {Makino}, J. 2001,
  \mnras, 321, 199

\bibitem[\protect\astroncite{{Sanders} \&
  {Lowinger}}{1972}]{1972AJ.....77..292S}
{Sanders}, R.~H., {Lowinger}, T. 1972, \aj, 77, 292

\bibitem[\protect\astroncite{Scalo}{1986}]{scalo86}
Scalo, J.~M. 1986, Fund. of Cosm. Phys., 11, 1

\bibitem[]{}
Spinnato, P., Fellhauer, M., Portegies Zwart, S.F., 2003, 
MNRAS in press (astro-ph/0212494)

\bibitem[\protect\astroncite{{Spitzer} \& {Hart}}{1971}]{1971ApJ...164..399S}
{Spitzer}, L.~J., {Hart}, M.~H. 1971, \apj, 164, 399

\bibitem[\protect\astroncite{{Tamblyn} \& {Rieke}}{1993}]{1993ApJ...414..573T}
{Tamblyn}, P., {Rieke}, G.~H. 1993, \apj, 414, 573

\bibitem[\protect\astroncite{{Testor} et~al.}{1993}]{1993A&A...280..426T}
{Testor}, G., {Schild}, H., {Lortet}, M.~C. 1993, \aap, 280, 426

\bibitem[\protect\astroncite{{Vishniac}}{1978}]{1978ApJ...223..986V}
{Vishniac}, E.~T. 1978, \apj, 223, 986

\end{thebibliography}
\end{document}